# Dual-dispersion-regime dual-comb mode-locked laser

## MACIEJ KOWALCZYK,[1,*] ŁUKASZ A. STERCZEWSKI,[1] XUZHAO ZHANG,[2,3] VALENTIN PETROV,[4] AND JAROSŁAW SOTOR[1]

[1]*Laser & Fiber Electronics Group, Faculty of Electronics, Photonics and Microsystems, Wroclaw University of Science and Technology, Wybrzeże Wyspiańskiego 27, 50-370 Wrocław, Poland*
[2]*State Key Laboratory of Crystal Materials, Shandong University, 250100 Jinan, China*
[3]*College of Information Science and Technology, Qingdao University of Science and Technology, Qingdao 266061, China*
[4]*Max Born Institute for Nonlinear Optics and Ultrafast Spectroscopy, Max-Born-Str. 2a, 12489 Berlin, Germany*
**Corresponding author: m.kowalczyk@pwr.edu.pl*

**We report on the first solid-state dual-comb mode-locked laser simultaneously operating in different dispersion regimes. Due to the intrinsic polarization-multiplexing in a birefringent Yb:$Ca_3NbGa_3Si_2O_{14}$ (Yb:CNGS) gain medium, the laser emits two cross-polarized pulse trains with a repetition rate offset of ~4.8 kHz from a single cavity. We obtain dual pulse generation with 20-fold difference in the duration, by setting the net cavity group delay dispersion to cross zero across the emission band of the employed gain medium. While the duration of the soliton-like pulses experiencing anomalous dispersion amounts to 117 fs, the second laser output that is spectrally located in the normal dispersion region, is strongly chirped with a pulse duration of 2360 fs.**

Locking two independent oscillators requires significant effort [1] compared to the intrinsic emission of mutually coherent beams from a shared laser cavity. Consequently, single-cavity dual-comb lasers (SCDCLs) [2] have emerged as interesting alternatives to pairs of independent optical frequency combs synchronized via electronic feedback loops [3]. Progress in the SCDCL field is primarily fueled by applications in dual-comb spectroscopy (DCS) [4], which enables broadband and high-resolution spectral measurements at high rates across different spectral regions. For most free-running SCDCL, the central DCS requirements of dissimilar repetition rates and high mutual comb coherence are naturally satisfied [5]. Therefore, demonstrations of demanding picometer-resolution measurements of high-finesse microcavities [6] or molecular transitions at low pressure [7] are routinely possible with these sources.

The first single-cavity dual-color femtosecond solid-state (Ti:sapphire) laser was demonstrated as early as 1993 [8]. At that time, however, asynchronous (due to different group delay dispersion; GDD) dual combs were not of particular interest because researchers focused on synchronizing the wavelength-multiplexed pulse trains. Therefore, their potential for use in DCS, or more generally, asynchronous optical sampling remained long unnoticed. This changed in recent years when numerous schemes for common-cavity dual-comb generation have been demonstrated. In particular, fiber-based bidirectional [9–13] and polarization-multiplexing [7,14,15] pulse shaping has greatly extended the dual-wavelength concept [6,16,17]. Although the most popular platform for realization of the SCDCL concept is still the optical fiber, solid-state (bulk) dual-comb lasers have also experienced remarkable growth [18–24]. Not only do they offer reduced nonlinearities permitting generation of high-power radiation [21,23], but also support ultrashort pulse duration [18] and exceptionally low-noise operation [24].

Whereas most studies have focused on the dual-comb generation mechanism, repetition rate tuning, and mutual stability characterization, some research groups have reported on diverse temporal and spectral patterns emerging in SCDCLs. These include nontrivial phenomena such as the bound soliton state [25] or the occurrence of asynchronous pulse trains forming in different dispersion regimes, which can be observed primarily in dual-wavelength SCDCLs. For instance, Wang *et al.* have reported the coexistence of a 830-fs-long stretched pulse with a 19.8-ps-long chirped (dissipative) soliton in a 1.95 μm Tm-doped fiber laser incorporating high third order dispersion [25]. Analogously-featured operation in a dispersion-managed Er-doped fiber laser was also presented [26]. Finally, a 1.05 μm dual-color Yb-doped fiber laser with mechanical dispersion tuning reported by Fellinger *et al.* enabled either generation of two solitons or operation in different dispersion regimes: normal for the blue-shifted and anomalous for the red-shifted pulses [17]. Notably, coexistence of temporally diverse pulses in *the same* dispersion regime has also been observed in fiber lasers generating two solitons with strongly differing spectral bandwidths [27,28].

Although these reports present the feasibility of realizing temporally and spectrally diverse dual-comb patterns in a shared laser cavity, all were implemented using optical fiber technology. This is because fiber lasers exhibit significantly higher intracavity nonlinearities and dispersion compared with their bulk counterparts, which greatly promotes exotic mode-locking regimes. In this letter, we address this niche and present the simultaneous generation of spectrally and temporally dissimilar pulses in a polarization-multiplexed solid-state Yb:CNGS dual-comb laser. Rather than ensuring anomalous dispersion for soliton mode-locking for the two combs, we promote the formation of cross-polarized pulses in both dispersion regimes (anomalous and normal), which enabled us to obtain a 20-fold difference in the pulse widths (117 fs vs. 2360 fs). Importantly for dual-comb applications, a slight improvement in the relative timing stability has also been observed when compared with the all-anomalous-dispersion scheme [24].

The schematic of the developed dual-comb laser cavity is shown in Fig. 1. As a gain medium we employ an antireflective (AR)-coated Yb(3%):CNGS crystal. This langasite-type calcium niobium gallium silicate offers a remarkably broadband emission spectrum that supports generation of sub-50 fs pulses, as demonstrated in our previous works [29,30]. The crystal used here has dimensions of 3×3×4 mm$^3$ and is cut with an angle $\theta$ = 36° between the normal propagation direction and the optical axis $c$. Both $c$ and one of the crystallographic $a$ axes of this trigonal crystal lie in the horizontal plane. The active element is pumped by a 1 W, single mode fiber-coupled laser diode operating at 979.4 nm. Due to the modest power delivered by the pump laser, no cooling of the active element is required. The pump waist within the crystal has a radius of 25 µm (measured) and matches well the estimated cavity laser mode radius of 22 µm (obtained using the ABCD matrix formalism). Stable mode-locking is initiated by a semiconductor saturable absorber mirror (SESAM) with a 500 fs recovery time and a 2.6% modulation depth. The output coupler has a transmission of 1%. The laser cavity is approximately 190 cm long which dictates the ~78.5 MHz pulse repetition rate.

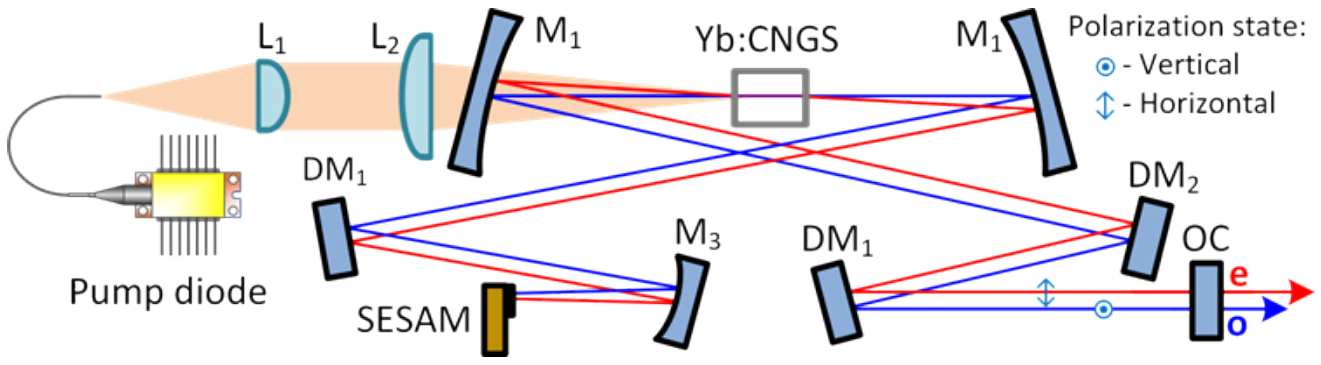

**Fig. 1.** Schematic of the dual-comb mode-locked Yb:CNGS laser. $L_1$ — 18.4 mm aspheric lens; $L_2$ — 100 mm spherical lens; $M_1$ — 100 mm concave spherical mirrors; $DM_{1,2}$ — dispersive mirrors; OC — output coupler; SESAM — semiconductor saturable absorber mirror; o/e beam — laser beams forming an ordinary/extraordinary ray inside the birefringent Yb:CNGS crystal.

One of the crucial aspects of the dual-dispersion laser design is clearly the estimation of the net cavity GDD. The major normal GDD contribution comes from the Yb:CNGS crystal introducing +660 fs$^2$ and +760 fs$^2$ per resonator round-trip at a 1050 nm wavelength, respectively for the beam polarized as an ordinary ($o$)- and an extraordinary ($e$; $\theta$ = 36°)-wave [31]. This is further increased by two curved folding mirrors $M_1$ with a total round-trip GDD equal to +400 fs$^2$ (simulation data provided by the manufacturer). Normal dispersion is counterbalanced by two types of dispersive mirrors ($DM_{1,2}$) with a total double-bounce GDD of −1200 fs$^2$ (measurement data). Figure 2a shows the round-trip GDD curves for these components along with that of the SESAM. Unfortunately, the measurement data of the latter were provided by the manufacturer only up to 1050 nm. No GDD data were available for $M_3$ and OC mirrors except that their dispersive contributions should be negligible.

The developed laser supports dual-comb generation via intrinsic polarization-multiplexing [24]. Due to the employed special cut of the birefringent Yb:CNGS gain medium, one can align the cavity to achieve simultaneous stable lasing of both $o$- and $e$-polarized beams. For that, the polarization of the pump radiation is set to form an $o$-ray inside the gain medium. Therefore, at normal incidence the pump and the $o$-polarized laser beam propagate collinearly inside the crystal. Once lasing of the $o$-beam is initiated, subsequent fine-alignment of the cavity end-mirrors allows one to trigger concurrent $e$-beam operation. Importantly, varying the mirrors' tilt provides full control over the power balance between the two generated beams, which is crucial regarding the Yb:CNGS gain anisotropy for σ ($o$-ray) and π ($e$-ray) polarizations [32]. After obtaining dual-polarization continuous wave operation, a stable mode-locked regime can be initiated with a mechanical perturbation of the SESAM.

Contrary to our previous work on a dual-comb laser emitting two soliton pulse trains [24], here we report simultaneous generation of pulses in two distinct dispersion regimes. Figure 2b presents the optical spectra together with the estimated round-trip net cavity GDDs experienced by the two beams. In this analysis, we excluded the SESAM contribution due to the limited spectral range of the available data. However, its dispersion profile (see Fig. 2a) signifies a further enhancement of the GDD contrast experienced by the two combs. The $o$-beam spectrum exhibits a peak at 1062 nm with a corresponding GDD of approx. −500 fs$^2$ (Fig. 2b). The spectrum resembles a soliton (sech$^2$-) shape with characteristic Kelly sidebands. In contrast, the spectrum of the $e$-beam exhibits a rectangular profile typical for normal dispersion regime [33]. Indeed, the peak wavelength of 1032 nm corresponds to a GDD of +400 fs$^2$.

The intensity autocorrelation measurement (Fig. 2c) yielded a pulse duration of 117 fs for the $o$-beam, which is relatively close to its Fourier limit (time-bandwidth product (TBP) of 0.44). On the contrary, the duration of the $e$-beam pulses amounted to 2.36 ps (TBP = 5.92), indicating the presence of strong chirp. This confirms that the two combs experienced completely different pulse formation mechanisms, despite being simultaneously emitted from the very same laser cavity. Nevertheless, they maintained dual-comb properties with a repetition rate offset ($\Delta f_{rep}$) of 4795 Hz, as shown in the radio-frequency (RF) spectrum (Fig. 2d). The additional low intensity peaks spaced by multiples of $\Delta f_{rep}$ arise due to parasitic intermodulation of the main beat notes in the photodetector with limited linearity.

Note that in the future works $\Delta f_{rep}$ can be tuned by placing an optical wedge into one of the intracavity beams. The essential laser output parameters are summarized in Table 1.

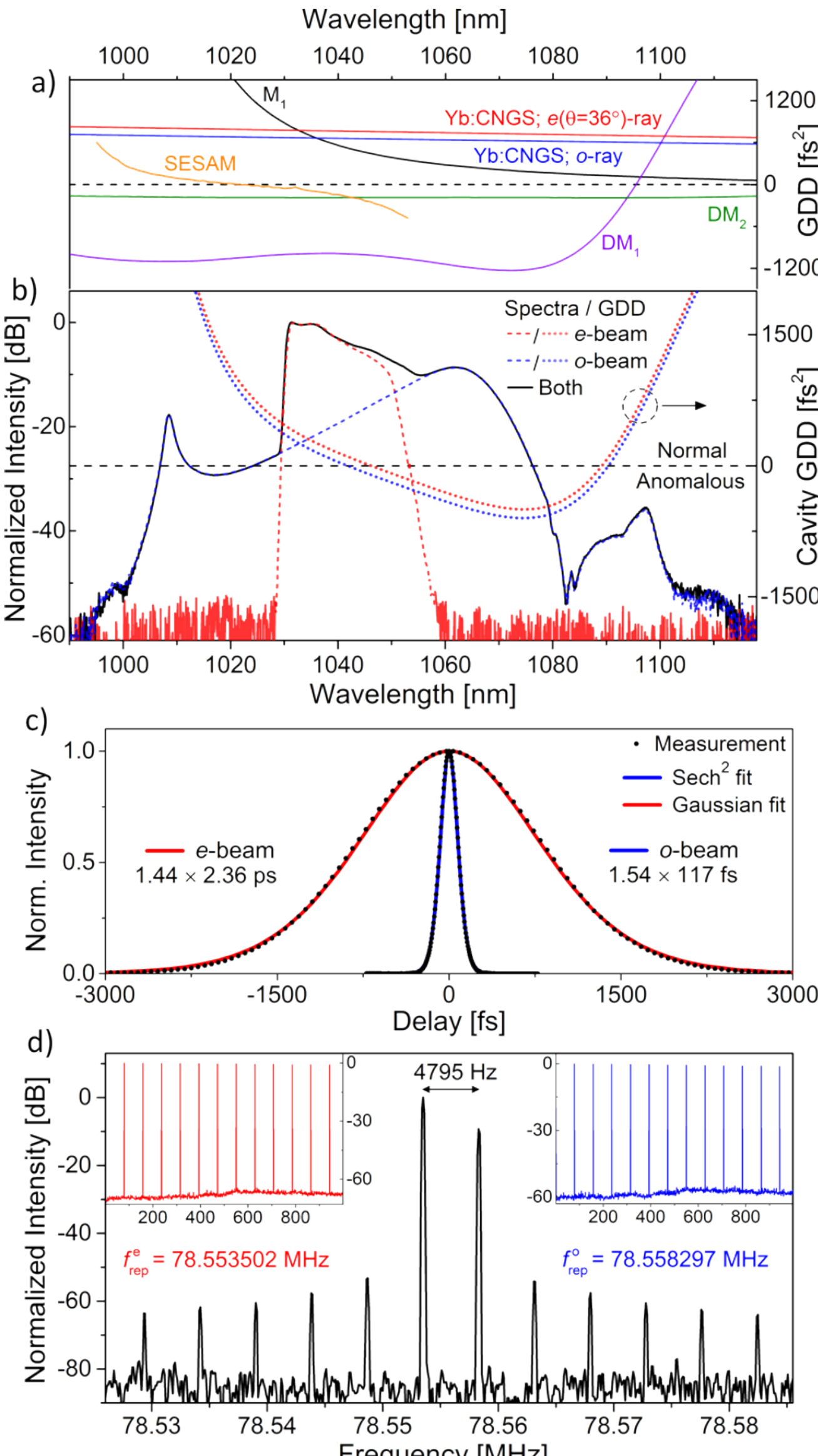

**Fig. 2.** (a) Round-trip GDD introduced by individual cavity components (two items of $M_1$ and $DM_1$ mirrors are implemented in the cavity and thus their contributions are doubled). Performance of the dual-comb mode-locked Yb:CNGS laser: (b) optical spectra of the *e*/*o*-beams and their superposition (logarithmic scale) together with the estimated round-trip net cavity GDD (excluding SESAM contribution); (c) corresponding autocorrelation traces with fits assuming sech$^2$- (*o*-beam) and Gaussian-shaped (*e*-beam) pulses; (d) corresponding RF spectra of the fundamental beat notes of both beams around 78.555 MHz (resolution bandwidth; RBW = 100 Hz); insets: 1-GHz-wide RF spectra of the individual beams (RBW = 1 kHz).

To further confirm that the laser simultaneously emits two cross-polarized beams rather than operates in a polarization-switching regime, we measured mutual beating between the generated combs in a simple dual-comb experiment. After separating the two beams by a polarizing beam-splitter, the polarization state of one of them is rotated with a half-wave plate for maximizing the interference signal. Subsequently, the beams are recombined with a 50:50 beam-splitter and pass through a home-built grating-based spectral bandpass filter. This enabled selection of narrowband radiation ($\Delta\lambda$ = 0.8 nm) to avoid aliasing of the down-converted comb. The two beams are then incident on a photodiode producing the beating signal, which is next amplified and low-pass filtered before acquiring with a fast oscilloscope in a similar way to that of ref. [24]. Figure 3 depicts an unprocessed trace of the registered electrical signal comprising a train of dual-comb interferograms with periodic bursts. Their temporal separation is defined by the inverse of the repetition rate offset $\Delta f_{rep}$ of 208 μs. The residual low-intensity spurious peaks visible between the consecutive bursts originate from parasitic intracavity reflections appearing due to imperfect AR-coatings of the gain crystal [24].

**Table 1.** Parameters of the output *e*/*o*-beams. $\lambda_0$ — peak emission wavelength; $\Delta\lambda$ — full-width-at-half-maximum (FWHM) spectral bandwidth; $\tau$ — FWHM pulse duration; TBP — time-bandwidth product; $P_{av}$ — average output power; $f_{rep}$ — repetition rate.

| | *e*-beam | *o*-beam |
|---|---|---|
| $\lambda_0$ | 1031.7 nm | 1061.7 nm |
| $\Delta\lambda$ | 8.9 nm | 14 nm |
| $\tau$ | 2360 fs | 117 fs |
| TBP | 5.92 | 0.44 |
| $P_{av}$ | 80 mW | 20 mW |
| $f_{rep}$ | 78.553502 MHz | 78.558297 MHz |

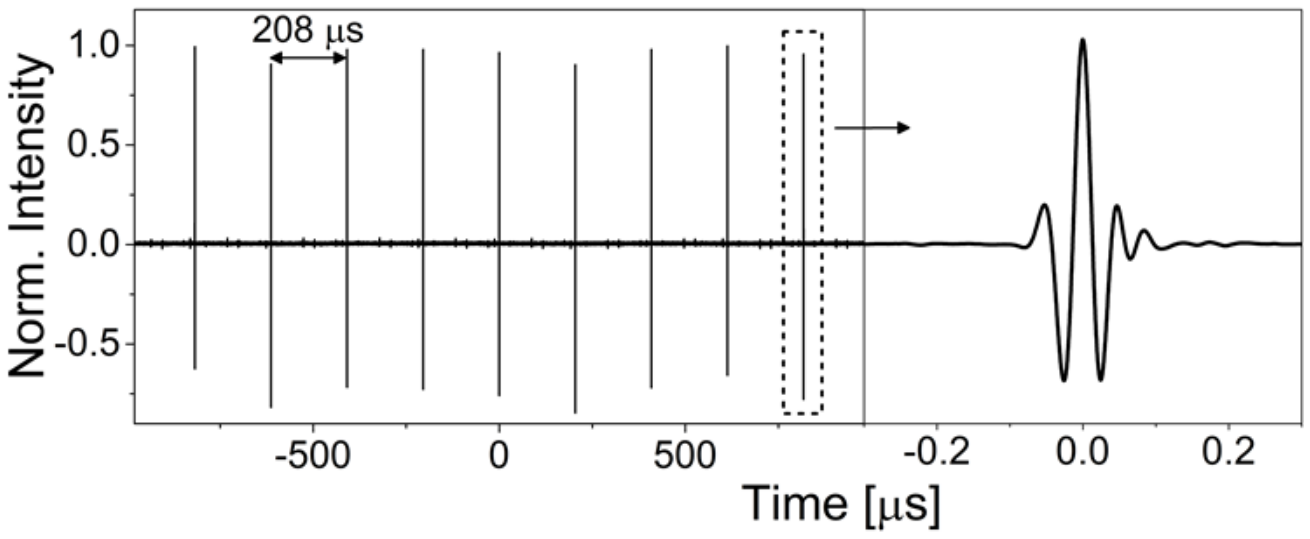

**Fig. 3.** A raw oscilloscope trace of the dual-comb interferogram with bursts spaced by $1/\Delta f_{rep}$ = 208 μs. One of the centerbursts is zoomed on the right.

For a more detailed comparison of the dual-dispersion dual-comb regime with the original all-anomalous-dispersion design [24] we studied their noise properties. Table 2 summarizes the most important parameters with the terminology adopted from ref. [34]: the relative RMS timing jitter between the two pulse trains ($\tau^{RMS}$), period timing jitter ($\tau^{period}$), and integrated phase noise of the relative carrier-envelope-offset frequency ($\int S_{\Delta fceo}$; see Supplemental Document for detailed explanation and additional data). We found out that the presented laser exhibits an improvement in repetition rate stability, while its relative $f_{ceo}$ noise remains virtually the same. We expect that this noise behavior may be attributed to the reduced spectral overlap between the combs around their central wavelengths (leading to a suppressed gain-competition-based cross-coupling) as well as a significantly lower peak power of the $e$-beam (which weakens nonlinear cross-coupling). Nevertheless, the exact explanation of this phenomenon requires further, more detailed investigations, which may be provided by pulse-to-pulse-resolved mode-locking dynamics studies [33,35], that will be covered by our future works.

**Table 2.** Comparison of relative stability performance between the all-anomalous [24] and dual-dispersion mode of operation of the Yb:CNGS dual-comb laser. $\tau^{RMS}$ — relative RMS timing jitter between the two pulse trains; $\tau^{period}$ — period timing jitter, $\int S_{\Delta fceo}$ — integrated phase noise of $\Delta f_{CEO}$ from $\Delta f_{rep}/2$ to 10 Hz (see Supplemental Document for details).

| | All-anomalous [24] | Dual-dispersion [this work] |
|---|---|---|
| $\tau^{RMS}$ | 20.7 ps | 2.5 ps |
| $\tau^{period}$ | 331.3 fs | 108.8 fs |
| $\int S_{\Delta fceo}$ | 9.7 rad | 11.2 rad |

In conclusion, we have demonstrated a mode-locked bulk solid-state laser simultaneously emitting two combs that exploits polarization-multiplexing in the Yb:CNGS gain medium. The net cavity GDD crossing zero across the laser emission band allowed generation of the individual combs in different dispersion regimes. Consequently, despite being simultaneously emitted from the very same laser cavity, the two orthogonally polarized pulse trains experienced distinct pulse formation mechanisms resulting in dissimilar temporal and spectral profiles. We achieved a 20-fold pulse duration difference with 117 fs and 2360 fs for the pulses shaped in the presence of anomalous and normal dispersion, respectively. The presented laser source opens a novel avenue for studying fundamental mode-locking dynamics by providing insight into interplay between pulses sharing the very same bulk laser cavity and gain medium, yet experiencing completely different pulse formation mechanisms [33, 35–37]. Moreover, we expect that the observed improvement of the relative timing jitter with respect to the dual-soliton regime [24] may help to develop more stable free-running single-cavity dual-comb lasers in the future.

**Funding.** National Science Centre (NCN, Poland) (2015/18/E/ST7/00296); Horizon 2020 Framework Programme (H2020) (101027721).

**Acknowledgments.** We would like to acknowledge Laser Zentrum Hannover for fabrication of the AR coatings on the Yb:CNGS sample. This project has received funding from the European Union's Horizon 2020 research and innovation programme under the Marie Skłodowska-Curie grant agreement No 101027721.

**Disclosures.** The authors declare no conflicts of interest.

**Data availability.** Data underlying the results presented in this letter may be obtained from the authors upon reasonable request. MATLAB scripts for comparing the relative offset frequency noise from dual-comb interferograms are openly available in figshare at: https://doi.org/10.6084/m9.figshare.25055684.v1.

**Supplemental document**. See Supplement 1 for supporting content.

## Full references

# Dual-dispersion-regime dual-comb mode-locked laser: supplemental document

## 1. NOISE CHARACTERISTICS

The different pulse shaping mechanisms of the dual-dispersion dual-comb laser presented in the primary document may influence the relative noise characteristics compared to the original all-anomalous-dispersion dual-comb laser design of ref. [1]. In particular, it is unclear how the two setups compare in terms of common phase noise suppression relevant for mode-resolved dual-comb spectroscopy. Because the relative timing and carrier-envelope offset (CEO) noise can be inferred from dual-comb interferograms, we processed 200-ms-long acquisitions using a computational phase retrieval and correction algorithm [2] to compare the two setups.

First we compare the timing stability, where the nomenclature of ref. [3] by Camenzind et al. is adopted. The relative root-mean-square (RMS) timing jitter $\tau^{\mathrm{RMS}}$ is calculated from the $n$-th interferogram arriving at $t_n$, and compared to the expected time-of-arrival (TOA) defined by multiples of the expected period $1/\left\langle \Delta f_{rep} \right\rangle$:

$$\tau^{\mathrm{RMS}}[n] = t_n - \frac{n}{\left\langle \Delta f_{rep} \right\rangle}. \tag{S1}$$

This quantity characterizes how much the delay between optical pulses from the two combs fluctuates compared to the jitter-free case. To retrieve the TOAs needed for the characterization, we employ a constant fraction discriminator [2], which provides a numerical trigger based on the digital difference frequency signal (DDFG) [4]. However, other techniques should serve this application equally well like detecting peaks of the dual-comb signal envelope [3] or finding the maximum of the cross-correlation/cross-ambiguity function [5,6]. Note that to relate the microwave and optical domains, we employ a comb factor $\Delta f_{\mathrm{rep}}/f_{\mathrm{rep}}$ in our calculations, as in ref. [3].

Another related quantity under study is the period timing jitter, which measures how much the duration of interferograms (difference of consecutive TOAs) fluctuates relative to the expected period. It is defined as

$$\tau^{\mathrm{period}}[n] = (t_n - t_{n-1}) - \frac{1}{\left\langle \Delta f_{\mathrm{rep}} \right\rangle}. \tag{S2}$$

Table S1 summarizes the numerical results obtained for the two lasers.

**Table S1.** Comparison between the all-anomalous and dual-dispersion mode of operation of the dual-comb mode-locked laser utilizing a Yb:CNGS gain medium.

| | Single-dispersion (all-anomalous) regime [1] | Dual-dispersion-regime |
|---|---|---|
| Relative RMS timing jitter between the two pulse trains | 20.7 ps | 2.5 ps |
| Period timing jitter | 331.3 fs | 108.8 fs |

Unexpectedly, we find that both the relative RMS timing jitter, and period timing jitter greatly improve when we employ the dual-dispersion setup. Consequently, the latter is better suited for free-running asynchronous optical sampling or dual-comb spectroscopy experiments. The exact origin of the noise performance improvement needs further research but we postulate that it may relate to (i) reduced spectral overlap between the combs; (ii) lower peak power of the multi-ps pulse that forms in the normal dispersion regime. When the two pulses temporally overlap inside the gain crystal, they compete for the gain. This often leads to severe pulse amplitude modulation, which manifests itself in polarization-multiplexed fiber lasers as slowly-varying post-centerburst oscillations in the dual-comb interferogram [7]. Here, the spectral overlap between the *o*- and *e*-beams around their respective peak wavelengths is significantly smaller when compared to ref. [1], which might have contributed to decreasing their gain-competition-based cross-coupling. Additionally, by lengthening the duration of one of the pulses to the multi-picosecond range, we expect to have greatly suppressed intensity-dependent inter-pulse nonlinear interaction.. Note that it is minor sub-Hz $\Delta f_{\mathrm{rep}}$ drifts that are mostly responsible for the higher jitter reported here, which correspond to a relative $f_{\mathrm{rep}}$ stability as low as $10^{-8}$. We also want to underline that the numbers quoted for the two free-running dual-comb laser setups are not a limitation *per se,* and can be greatly improved with a more robust mechanical cavity design [3].

Equally important for performance evaluation is the phase of the relative carrier-envelope-offset (CEO) frequency ($\Delta f_{\mathrm{CEO}}$), or simply the interferogram carrier-envelope phase $\Delta\varphi_0$. For a stable dual-comb laser with $\Delta f_{\mathrm{CEO}} \neq 0$, $\Delta\varphi_0$ is expected to evolve linearly. Unfortunately, a free-running non-CEO-stabilized dual-comb laser exhibits highly nonlinear phase excursions instead. In Fig. S1, we characterize the power spectral density (PSD) of computationally-retrieved $\Delta\varphi_0$ compensated for the linear trend. The top panel illustrates the frequency spectrum (power spectral density) of phase excursions, while the bottom shows the frequency-integrated phase noise retrieved from the PSD ($\int S_{\Delta fceo}$). Note that $\Delta\varphi_0$ is sampled only around interferogram centerbursts with a sampling frequency of $\Delta f_{\mathrm{rep}}$, which limits the analysis frequency range to $\Delta f_{\mathrm{rep}}/2$.

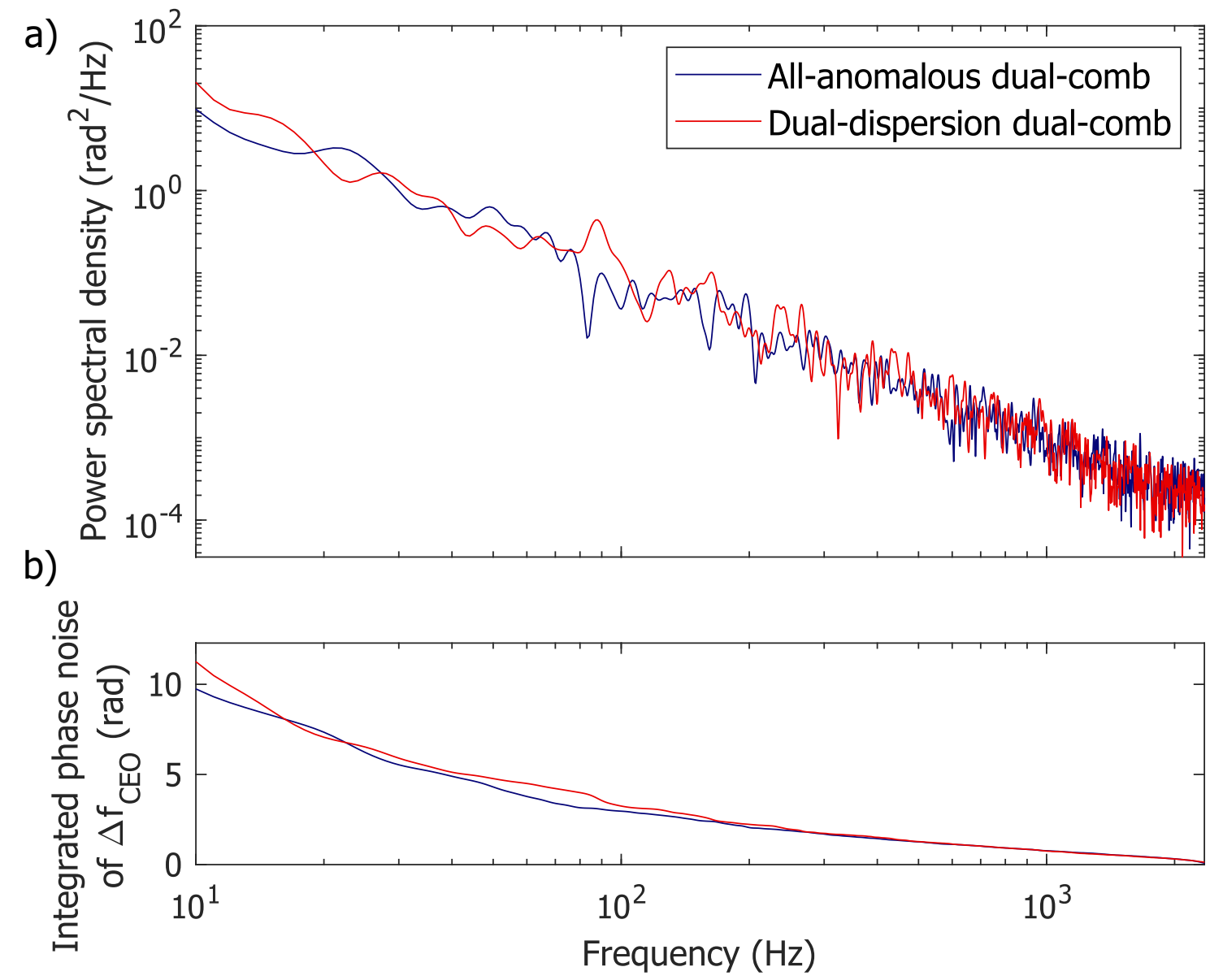

**Fig. S1.** Relative offset frequency noise characterization retrieved from dual-comb interferograms for the anomalous-only dispersion regime laser [1], and the dual-dispersion regime laser discussed in the main manuscript. (a) Power spectral density (PSD) of relative CEO phase fluctuations. (b) Integrated phase noise retrieved from the PSD.

Unlike in the timing jitter case, we do not see significant differences between the two laser configurations. Both display nearly the same relative carrier-envelope phase noise characteristics with a dominant $1/f^2$ slope (white frequency noise rolling off −20 dB/decade) and only minor local differences occurring at acoustic/mechanical noise frequencies. This further confirms the importance of stable mechanical cavity design for low-timing-jitter and low-phase-noise operation, which may be potentially degraded by environmental noise effects.